\documentclass[useAMS,usenatbib]{mn2e}

\usepackage{graphicx}

\def\aap{{A\&A}}		
 
\def\apj{{ApJ}}
\def\apjl{{ApJ}}		
\def\apjs{{ApJS}}		
\def\aj{{AJ}}
 
\def\mnras{{MNRAS}}
\def\msun{M_\odot}

\title[Monte Carlo Simulations of Star Clusters - V. The globular
  cluster M4]{Monte Carlo Simulations of Star Clusters -
  V.  The globular cluster M4}
\author[M. Giersz and D.C. Heggie]{Mirek Giersz$^{1}$\thanks{E-mail:
mig@camk.edu.pl (MG); d.c.heggie@ed.ac.uk (DCH)} and Douglas C.
Heggie$^{2}$
\\
$^{1}$Nicolaus Copernicus Astronomical Centre, Polish Academy of Sciences, ul. Bartycka 18, 00-716 Warsaw, Poland\\
$^{2}$University of Edinburgh, School of Mathematics and Maxwell
Institute for Mathematical Sciences, King's
Buildings, Edinburgh EH9 3JZ, UK}
\begin{document}

\date{Accepted \ldots. Received \ldots; in original form \ldots}

\pagerange{\pageref{firstpage}--\pageref{lastpage}} \pubyear{2002}

\maketitle

\label{firstpage}

\begin{abstract}
We describe Monte Carlo models for the dynamical evolution of the
nearby globular cluster M4.  The code includes treatments of two-body
relaxation, three- and four-body interactions involving primordial
binaries and those formed dynamically, the Galactic tide, and the
internal evolution of both single and binary stars.  We arrive at a
set of initial parameters for the cluster which, after 12Gyr of
evolution, gives a model with a satisfactory match to the surface
brightness profile, the velocity dispersion profile, and the
luminosity function in two fields.  We describe in particular the
evolution of the core, and find that M4 (which has a classic King
profile) is actually a {\sl post-collapse} cluster, its core radius
being sustained by binary burning.  We also consider the distribution
of its binaries, including those which would be observed as
photometric binaries and as radial-velocity binaries.  We also
consider the populations of white dwarfs, neutron stars, black holes
and blue stragglers, though not all channels for blue straggler
formation are represented yet in our simulations.
\end{abstract}


\section{Introduction}

The present paper opens up a new road in the study of the dynamical
evolution of globular clusters.  We adopt the Monte Carlo method of
Giersz \citep{Gi1998,Gi2001,Gi2006}, which in recent years has been enhanced to
deal quite realistically with the stellar evolution of single and
binary stars, to study the dynamical history of the nearby globular
cluster M4.  An earlier version of the code had already been used to
study the dynamical history of $\omega$ Cen \citep{GH2003}, but at
that time the treatment of stellar evolution was primitive and there
were no binaries.  The new code has been thoroughly tested on smaller
systems, by comparison with $N$-body simulations and observations of
the old open star cluster M67 \citep{GH2008}.  There we showed that
the Monte Carlo code could produce data of a similar level of detail
and realism as the best $N$-body codes.  Now for the first time we
consider much richer systems, with about half a million stars
initially, which are at present beyond the reach of  $N$-body methods.

This paper has a place within a long tradition of the modelling of
globular star clusters, but the place is a distinctive one.  First, we
are not concerned with a static model of a star cluster at the present
day, like a King model.  We are concerned with issues where the
dynamical history of the star cluster is important, where static
models are uninformative.  Secondly, our aim is to construct a model
of a specific star cluster, rather than trying to understand the
general properties of the evolution of a population of star clusters.
This has been done before, and a brief history is outlined in
\citet{GH2008}, but the present work takes these efforts onto a new
level of realism, in terms of the description of stellar evolution,
and dynamical interactions involving binary stars.

This problem is not easy.  Not only is it necessary to use an
elaborate technique for simulating the relevant astrophysical
processes, but it is necessary also to search for initial conditions
which, after about 12Gyr of evolution, lead to an object resembling a
given star cluster.  By ``resembling'' we do not simply mean matching
the overall mass, radius and binary fraction of a cluster, for example, for two
reasons:\begin{enumerate}
\item  We have found that values found for data in the
literature are highly uncertain, and different sources are
contradictory.  These data are usually derived, in some
model-dependent way, from such data as surface-brightness profiles and
velocity dispersion profiles, and we prefer to compare our models
directly with this data, and not with inferred global parameters.
\item  We have found that, even if one achieves a satisfactory fit to
these profiles, the model may give a very poor comparison with the
luminosity function.  
\end{enumerate}
From these considerations we conclude that a model which aims to fit
only the mass and radius of a star cluster (say) may be very far from the
truth.  

Tackling this difficult problem is not just interesting, however.  We
have been motivated by a number of pressing astrophysical problems.
For example, the two nearby star clusters M4 (the subject of this
study) and NGC 6397 (which we shall consider in our next paper in this
series) have rather similar mass and radius, and yet one has a classic
King profile, while the other is a well-studied example of a cluster
with a ``collapsed core'' \citep{Tr1995}.  Among possible
explanations one may consider differences in the population of
binaries, which are known to affect core properties, or in tidal
effects.  Indeed the present paper will show that these two clusters
may be much more similar than one would suppose from the surface
brightness profiles alone.  

A second motivation for our work is our involvement in observational
programmes aimed at characterising the binary populations in globular
clusters.  What differences (e.g. in the
distributions of periods and abundances) should one expect to find
between the core and the halo?  These issues are important in the
planning of observations, and in their interpretation.

The cluster M4 is the focus of much of this effort because it is
nearby, making it a relatively  easy target for deep observational
study.  It was the first globular cluster to yield a deep sequence of
white dwarfs \citep{Ri1995}.  More recently it has been subjected to an
intensive observational programme by the Padova group \citep{bedinetal2001,bedinetal2003,andersonetal2006}, which includes searches for radial-velocity binaries in the
upper main sequence \citep{Somm}.
It also turns out to be a cluster which (we conclude)
started with only about half a million stars, which facilitates the
modelling.  Along with the open cluster M67, M4 was chosen by the
international MODEST consortium, at a meeting in Hamilton in 2005, as
the focus for joint effort by theorists and observers, to cast light
on its binary population and dynamical properties.  M67 has been
modelled very successfully by \citet{Hu2005}, using $N$-body
techniques, and this paper represent the first theoretical step in a
similar study of M4.

The paper is organised as follows.  First, we summarise features of
the code and the models, the data we used, and our approach to the problem of finding
initial conditions for M4.  Then we present data for our best models:
surface brightness and velocity dispersion profiles, luminosity
functions, the properties of the binary population, white dwarfs and
other degenerate remnants, and the inferred dynamical state of the
cluster.  The final section summarises our findings and discusses them
in the context of work on other clusters, including objects to which
we will turn in future papers.

\section{Methods}

\subsection{The Monte Carlo Code}

The details of our simulation method have been amply described in
previous papers in this series.  Each star in a spherical star cluster
is represented by its mass, energy and angular momentum, and its
stellar evolutionary state may be computed at any time using synthetic
formulae for single and binary evolution.  It may be a binary or a
special kind of single star that has been created in a collision or
merger event.

Neighbouring stars interact with each other in accordance (in a
statistical sense) with the theory of two-body relaxation.  If one or
both of the participants is a binary, the probability of an encounter
affecting the internal dynamics is calculated according to analytic
cross sections, which also determine the outcome.  This is one of the
main shortcomings of the code, as these cross sections are not well
known in the case of unequal masses, and also the possibility of
stellar collisions during long-lived temporary capture is excluded.

A star or binary may escape if its energy exceeds a certain value,
which we choose to be lower than the energy at the nominal tidal
radius, in order to improve the scaling of the lifetime with $N$, as
explained in \citet{GH2008}.  This is the second main shortcoming of
the models, as it leads to a cutoff radius of the model that is
smaller than the true tidal radius, and this lowers the surface
density profile in the outer parts of the system.  

A difficulty in applying the Monte Carlo code to M4 is that it employs
a static tide, whereas the  orbit of M4 appears to be very elliptical
\citep{Di1999}.  We have to assume that a cluster can be placed in a
steady tide of such a strength that the cluster loses mass at the same
average rate as it would on its true orbit.  Some support for this
procedure comes from $N$-body modelling.  \citet{BM2003} show that
clusters on an elliptical orbit between about 2.8 and 8.5kpc dissolve
on a time scale intermediate between that for circular orbits at these
two radii, and that the dissolution time scales in almost the same way
with the size of the system.  \citet{wilkinsonetal2003} show that the
core radius of a cluster on an elliptic orbit evolves in very nearly
the same way as in a cluster with a circular orbit at the
time-averaged galactocentric distance.

  All other free
parameters of the code (e.g. the coefficient of $N$ in the Coulomb
logarithm) take the optimal values found in the above
study. 

\subsection{Initial Conditions} 

The initial models are as specified in Table \ref{tab:ics}.  Many of
these features (e.g. the properties of the binaries, except for their
overall abundance) were inherited from our modelling of the open
cluster M67, and those were in turn mainly drawn from the work of
\citet{Hu2005}.  Some of the parameters were taken to be freely
adjustable, and this freedom was exploited in the search for an
acceptable fit to the current observational data.

\begin{table}
\begin{center}
\caption{Initial parameters for M4}
\begin{tabular}{ll}
\hline
Fixed parameters\\
\hline
Structure&Plummer model\\
Initial mass function$^1${}&\citet{Kr1993} in the\\
& range [0.1,50]$\msun$\\
Binary mass distribution	&\citet{Kr1991}\\
Binary mass ratio               &Uniform (with component\\
&masses restricted  as for \\
&single stars)\\
Binary semi-major axis          &Uniform in log, $2(R_1+R_2)$\\
& to 50AU\\
Binary eccentricity             &Thermal, with eigenevolution \\
&\citep{Kr1995}\\
Metallicity $Z$&0.002\\
Age&12Gyr \citep{Ha2004}\\
\hline
Free parameters\\
\hline
Mass			&$M$\\
Tidal radius&$r_t$\\
Half-mass radius&$r_h$\\
Binary fraction 	&	$f_b$\\
Slope of the lower mass\\
 function&$\alpha$ (Kroupa = 1.3)\\
\hline
\end{tabular}
\label{tab:ics}

\end{center}
  \end{table}

\subsection{Observational data and its
  computation}\label{sec:data_computation}

Our first task was to iterate on the initial parameters of our model
in order to produce a satisfactory fit to a range of observational
data at an age of 12Gyr.  The data we adopted are as follows:
\begin{enumerate}
  \item Surface brightness profile:  here we used the compilation by
  \citet{Tr1995}, where the surface brightness is expressed in $V$
  magnitudes per square arcsec.   
\item Radial velocity profile:  this came from \citep{Pe1995}, and is
  the result of binning data the on the radial velocities of nearly
  200 stars.  Strictly we should refer to the line-of-sight velocity
  dispersion, as ``radial'' velocity has a different meaning in the
  Monte Carlo model.
\item The V-luminosity function (\citealt{Ri2004}, from which we
  considered the results for the innermost and outermost of their four
  annuli).  These data are lack correction for completeness, though
  the completeness factor for the outermost field is plotted in
  \citet{richeretal2002}.  For main sequence stars it is almost 100\%
  down to $V=15$, and drops steadily to less than 50\% at $V=17$.  For
  the innermost field the completeness correction would be larger.
\end{enumerate}

Now we consider how to compare this data with the output of the Monte
Carlo code.  This includes a list of each particle in the simulation,
along with data on its radius, radial and transverse velocities and
absolute magnitude, and numerous other quantities.  To construct the
surface brightness, we think of each particle as representing a
luminous shell of the same radius, and superpose the surface
brightness of all shells.  While this procedure involves a minimum of
effects from binning, or randomly assigning the full position of the
star, a shell has an infinite surface brightness at its projected
edge.  The effect of this, especially from the brightest stars, will
be seen in some of the profiles to be presented in this paper.
A similar problem arises in actual observations, and is often handled
by simply removing the brightest stars from the surface brightness
data presented.  The output from the model is corrected for extinction
(Table \ref{tab:m4dat}.  The distance of the line of sight from the
cluster centre is converted between pc (as in the model) and arcsec
(as in the observations) using the distance in the same table.

The line-of-sight velocity dispersion is computed in a similar way.
For each particle we calculate the mean square line-of-sight velocity
(because the orientation of the transverse component of the velocity
is random), and then sum over all particles.   In this sum, each
particle is weighted by a geometrical factor proportional to the
surface density of the particle's shell along the line of sight.  The
result is an estimate of the velocity dispersion that is weighted by
neither mass nor brightness, but only number.   To check the effects
of this, we have sometimes calculated velocity dispersions with
various cutoffs in the magnitudes of the stars included.  The effects
are usually quite small.

Computation of the luminosity function is the least problematic.  We
count stars in each bin in $V$-magnitude, but lying above a line in
the colour-magnitude diagram just below the main sequence.  Again the
contribution of each star is weighted by the same geometrical factor,
and the $V$ magnitude is corrected for extinction.

\begin{table}
\begin{center}
\caption{Properties of M4}
\begin{tabular}{lll}
\hline
Distance from sun	&	$^a$1.72kpc&\\
Distance from GC	&	5.9kpc&\\
Mass			&	$^a$63 000$\msun$&\\
Core radius		&	0.53 pc&\\
Half-light radius	&	2.3 pc&\\
Tidal radius		&	21 pc&\\
Half-mass relaxation time ($R_h$)	&660 Myr&\\
Binary fraction 	&	$^a$1-15\%&\\
$\left[\right.$Fe/H$]$&-1.2&\\
Age			&$^b$12Gyr&\\
$A_V$ 			&$^a$1.33&\\
\hline
\end{tabular}\label{tab:m4dat}
\end{center}
References: All data are from the current version of the catalogue of \citet{Ha1996}, except $^a$ \citet{Ri2004} (though this is not always the original reference for the quoted
number) and $^b$ \citet{Ha2004}.
  \end{table}

\subsection{Finding initial conditions}

Here we summarise our experience in approaching this problem.  A
number of studies (e.g. \citet{BM2003}, \citet{La2005}) give simple
formulae for the evolution with time of the bound mass of a rich star
cluster.  It is possible to derive similar simple formulae for the
evolution of the half-mass radius and other quantities.  There are
several problems with inverting these formulae, however, i.e. using
them to infer the initial parameters of a cluster from its present
mass and radius.  First, these present-day global parameters are quite
uncertain, even to within a factor of two.  Second, these formulae
depend on the galactic orbit and other parameters which are equally
subject to uncertainty.  Therefore we have adopted the more
straightforward but more laborious approach of iterating on the initial
parameters of our models (Tab \ref{tab:ics}, lower half); that is, we
select values for the five stated parameters, run the model, find
where the match with the observations is poor, adjust the parameters,
and repeat cyclically.

We have employed two methods to facilitate this process to some
extent.  First, we have often carried out mini-surveys, i.e. small,
coarse grid-searches around a given starting model, to find out how
changes in individual parameters affect the results.  Second, we have
used scaling to accelerate the process, and we now describe this
method in a little detail.

Suppose we wish to represent a star cluster which has mass $M$ and
radius $R$ with a model representing a cluster with a (usually
smaller) mass $M\ast$ and radius $R\ast$.  Since two-body relaxation
dominates much of the dynamics, we insist that the two clusters have
the same relaxation time, and so
$$
\frac{R\ast}{R} =
\left(\frac{M}{M\ast}\right)^{1/3}\left(\frac{\log\gamma
  N\ast}{\log\gamma N}\right)^{2/3},
$$
where $N,N\ast$ are the corresponding particle numbers.  Thus the
model of lower mass has larger radius.   The tidal radius is scaled in
the same way.
Then the observational
results (surface brightness profile, etc) can be computed for the
model of lower mass and then rescaled (by appropriate factors of the mass and
radius) to give a result for the more massive cluster (assuming that
the evolution is dominated by the processes of relaxation, stellar
evolution and tidal stripping).  In fact we have found that this is
very successful, in the sense that the inferred best values for the
initial conditions change little when runs are carried out with the
``correct'' (i.e. unscaled) initial mass, certainly when the
proportion of binaries is 10\% or less.  Some aspects of the evolution
are not well described by this scaling technique.  For example, we do
not change the distribution of the semi-major axes of the binaries.
In this way the internal evolution of the binaries is correctly
modelled (provided that the binaries remain isolated dynamically),
though their dynamical interactions with the rest of the system are
not.  In principle one could scale the semi-major axes in the same way
as $R$, but then the internal evolution of the binaries would be
altered.

\section{Models of M4}

\subsection{Finding initial parameters}

Our starting point was our work on the old open cluster M67
\citep{GH2008}, but with a larger initial mass and radius.
(\citet{BM2003}, for example, suggested that the initial mass of M4
(NGC 6121) was of order $7.5\times10^5\msun$, though we used
somewhat smaller values.)  To begin with, our choice of initial tidal
radius was inferred from the initial mass and the present-day
estimates of mass and tidal radius given in Table \ref{tab:m4dat},
assuming that $r_t\propto M^{1/3}$.  At first we adopted similar
values for the ``concentration'' ($r_t/r_h$) and binary fraction as in
the modelling of M67, but found that the surface brightness profile fitted poorly (with too large a core) unless the binary fraction was much smaller (5 to 10\%) and the concentration much higher.

Before describing our best models, it is worth briefly mentioning one
which provided a satisfactory fit to the surface brightness profiles.
The fit to the velocity dispersion profile was tolerable, but
indicated a model that was too massive by a factor of about 1.4.  Its
main flaw, however, was in the luminosity function, which was
generally too large by a factor in the range 2--3.  There are two
reasons why this is interesting.  One is that, for a long time, models
of star clusters were constructed entirely on the basis of the surface
brightness and velocity dispersion profiles.  It should be realised
that such models may be misleading in other ways.  Second, this
experience underscores the importance of properly normalised
luminosity functions.  In other words, it is important to know the
area of the field where the stars have been counted, or some
equivalent representation of properly normalised data.  Very often,
the emphasis is solely on the {\sl shape} of the luminosity function,
but we have found that the absolute normalisation is an essential
constraint.

\citet{BM2003} show that the lower mass function becomes flatter as
the fraction of mass lost by the cluster increases (i.e. towards the
end of its life).  Therefore we could perhaps have improved the fit
with the luminosity function by starting with a more massive model and
somehow ensuring a larger escape rate so as to leave a similar mass at
the present day.   Instead, we elected to change the slope of the
low-mass IMF from the canonical value of $\alpha=1.3$ \citep{Kr2007a}
to $\alpha=0.9$.  (There is some justification for a lower value
for low-metallicity populations, though it has been argued
\citep{Kr2007b} that there is no pristine low-metallicity population
where the IMF can be inferred securely.)  


\subsection{A Monte Carlo model of M4}

By some
experimentation we arrived at a model which gave a fair fit to all
three kinds of observational data; see Table 3 (where we compare with
a King model developed by \citet{Ri2004}, and Figs 1-4. 
It is worth noting that no arbitrary normalisation has
been applied in these comparisons between our model and the
observations.  The surface brightness profile, for example, is
computed directly from the V-magnitudes of the stars in the Monte
Carlo simulation, as described in Sec.\ref{sec:data_computation}.  In
the construction of a King model, by contrast, it is often assumed
that the mass-to-light ratio is arbitrary.

\begin{table*}
\begin{minipage}[]{80mm}
 \begin{center}
\caption{Monte Carlo and King models for M4}
\begin{tabular}{llll}
\hline
Quantity&MC model &MC model &King model\\
&($t = 0$)&($t=12$Gyr)& \citep{Ri2004}\\
\hline
Mass ($\msun$)&$3.40\times10^5$&$4.61\times10^4$&\\
Luminosity ($L_\odot$)&$6.1\times10^6$&$2.55\times10^4$&$6.25\times10^4$\\
Binary fraction $f_b$&0.07&0.057&0\\
Low-mass MF slope $\alpha$&0.9&0.03&0.1\\
Mass of white dwarfs ($\msun$)&0&$1.81\times10^4$&$3.25\times10^{4^\ast}$\\
Mass of neutron stars ($\msun$)&0&$3.24\times10^3$\\
Tidal radius $r_t$ (pc)&35.0&18.0&\\
Half-mass radius $r_h$ (pc)&0.58&2.89&\\
\hline
\end{tabular}\label{tab:mc_king}
\end{center}
$\ast$: this is the quoted mass of ``degenerates''
\end{minipage}
\end{table*}

  \begin{figure}
  \begin{minipage}[]{0.45\textwidth}
{\includegraphics[height=10cm,angle=-90,width=10cm]{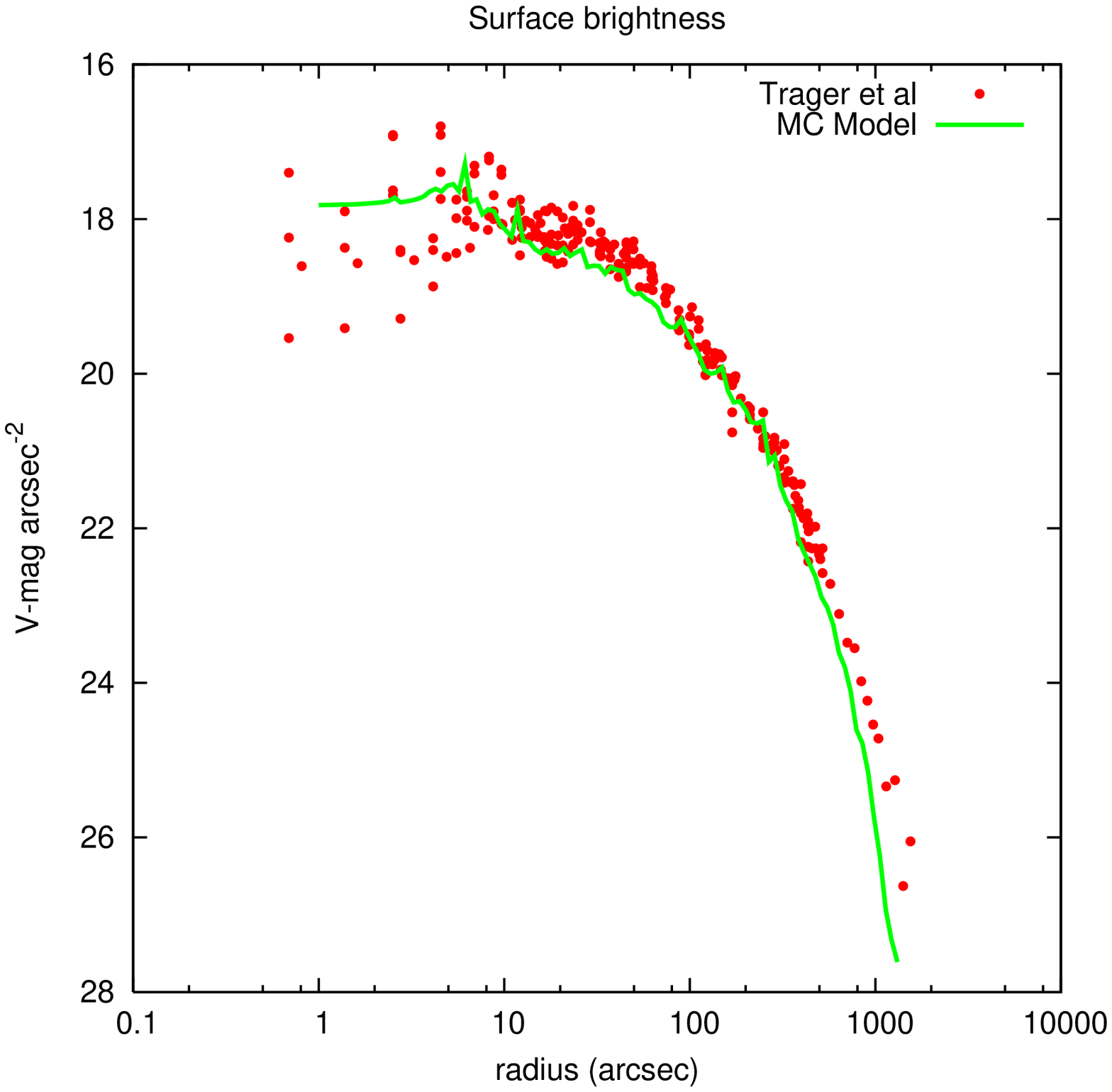}}
    \caption{Surface brightness profile of our Monte Carlo model,
    compared with the data of \citet{Tr1995}.}\label{fig:sbp}
\end{minipage}
\hfill
  \begin{minipage}[]{0.45\textwidth}
{\includegraphics[height=10cm,angle=-90,width=10cm]{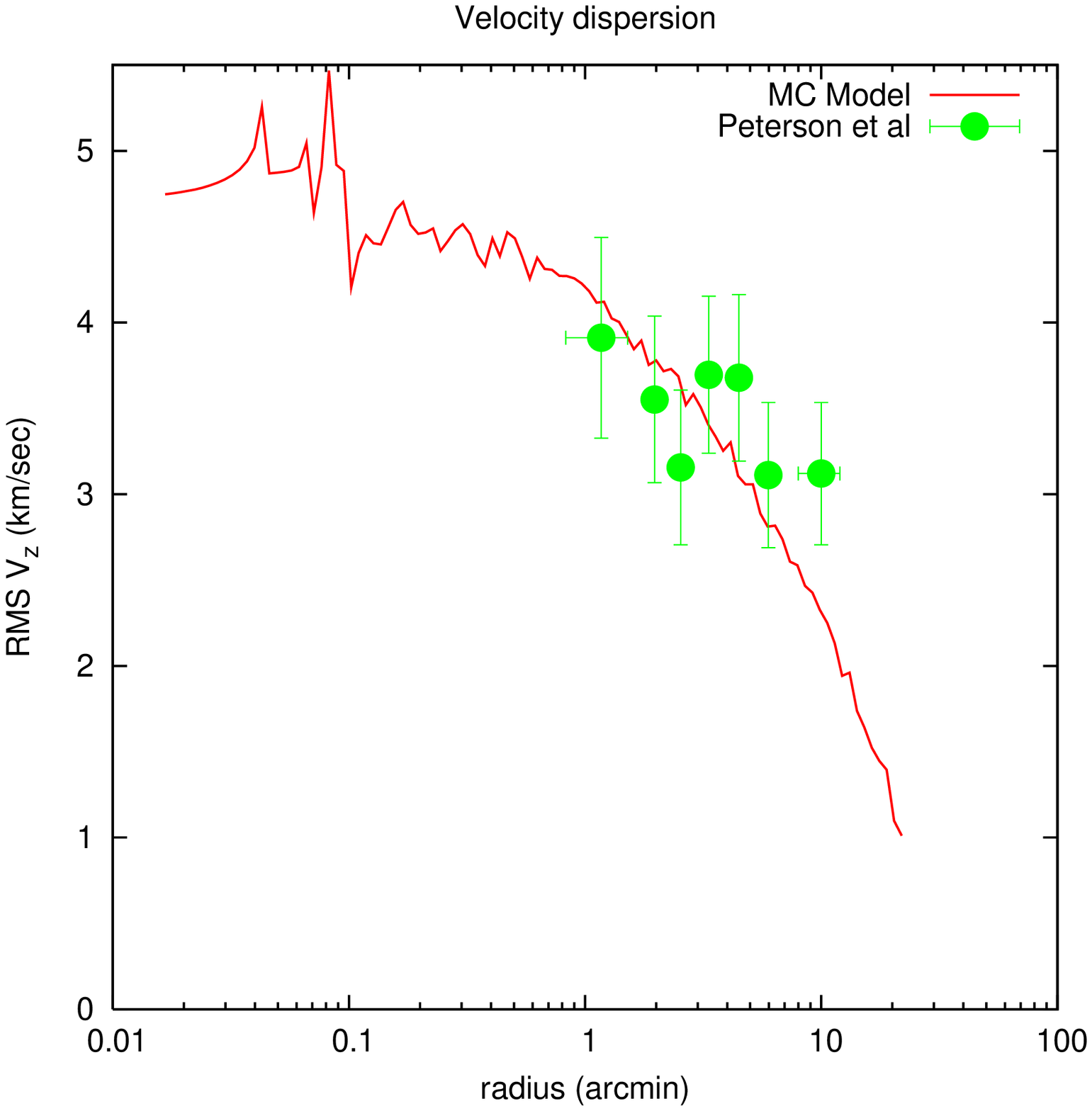}}
    \caption{Velocity dispersion  profile of our Monte Carlo model,
    compared with the data of \citet{Pe1995}.}\label{fig:vdp}
  \end{minipage}
\end{figure}

  \begin{figure}
    \begin{minipage}[]{0.45\textwidth}
{\includegraphics[height=10cm,angle=-90,width=10cm]{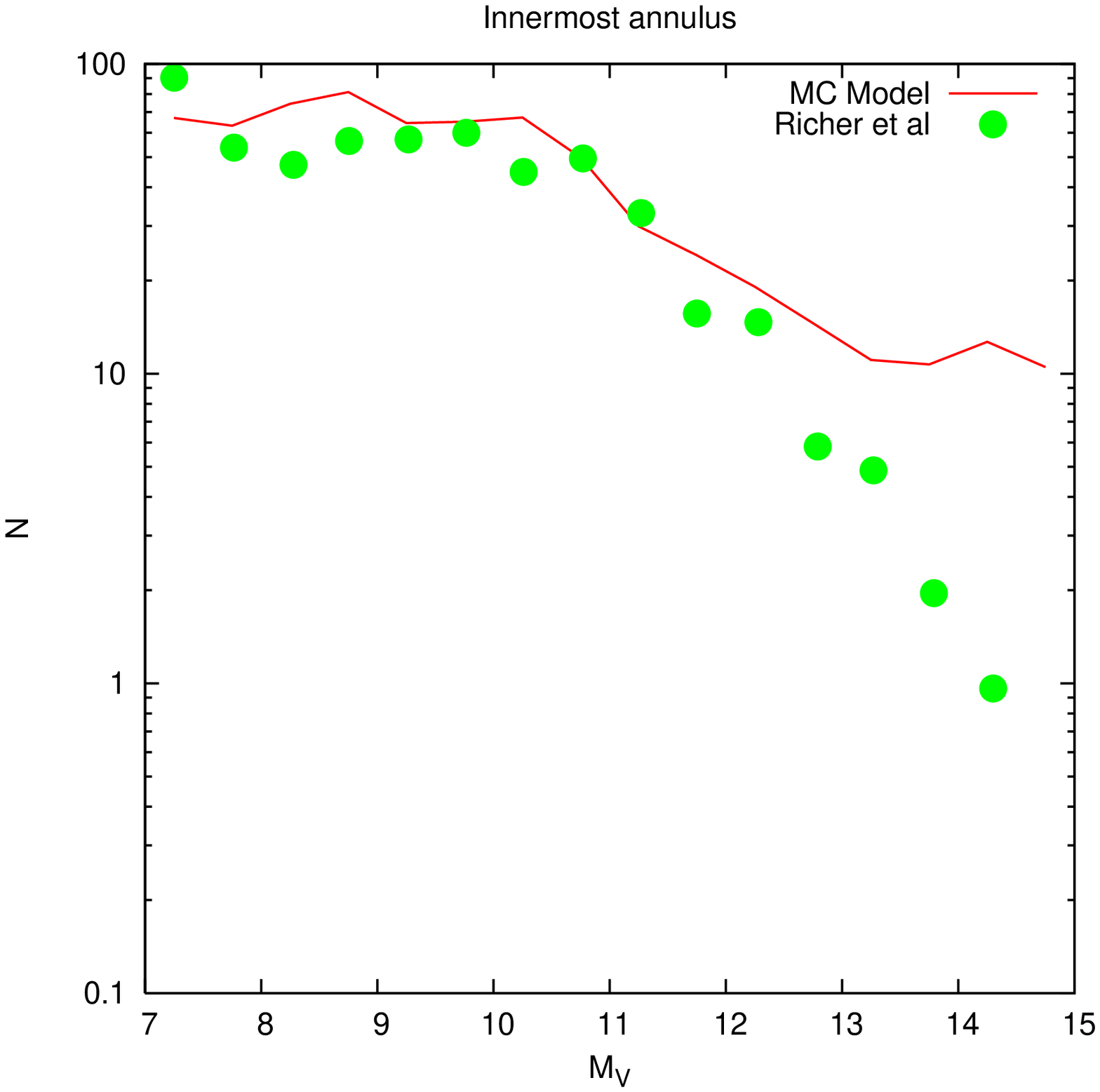}}
    \caption{Luminosity function  of our Monte Carlo model at the
    median radius of the innermost annulus in \citet{Ri2004}, compared with their data.}
    \end{minipage}
\hfill
    \begin{minipage}[]{0.45\textwidth}
{\includegraphics[height=10cm,angle=-90,width=10cm]{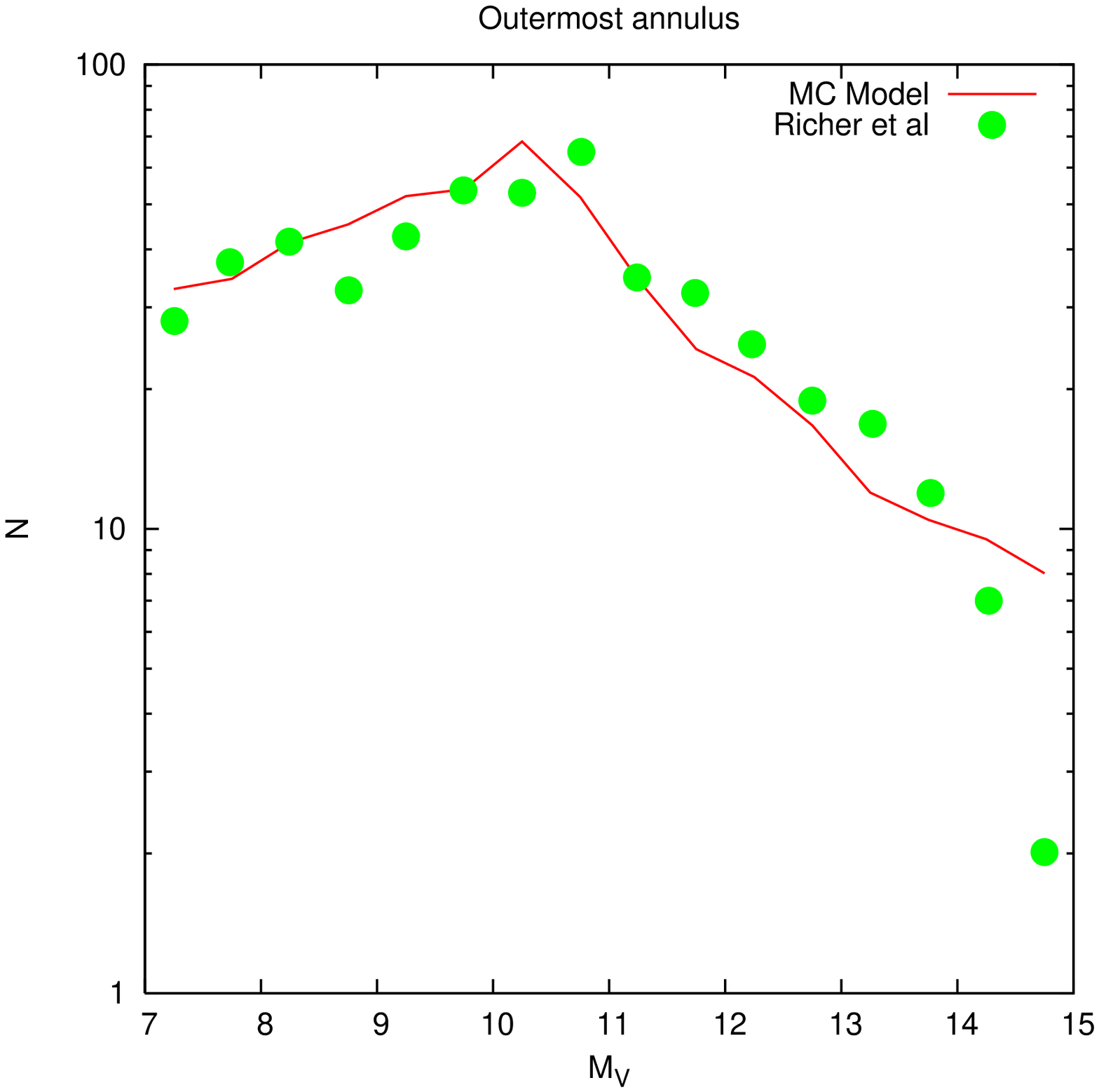}}
    \caption{Luminosity function  of our Monte Carlo model at the
    median radius of the outermost annulus in \citet{Ri2004}, compared with their data.}
    \end{minipage}
  \end{figure}

\subsubsection{Surface brightness}

While the overall surface brightness profile is slightly faint, the
most noticeable feature of Fig.\ref{fig:sbp} is that the model has a
somewhat smaller limiting radius than the observational data.  The
reason for this is explained in \citet{GH2008}: in short, we impose a
smaller tidal radius than the nominal tidal radius, in an
$N$-dependent way, which is intended to ensure that the overall
rate of escape from the model behaves in the same way as in an
$N$-body simulation.  Another point to notice is the disagreement
between the total luminosity of our model and that of the King model quoted in the final column
of Table \ref{tab:mc_king}.  \citet{Tr1995} give an analytic fit to the
surface brightness profile, and we have checked that the integrated
value is close to ours.

The data for the Monte Carlo model in Table \ref{tab:mc_king} would be
consistent with a galactocentric radius of about 1.7kpc, in an
isothermal galaxy model with a circular velocity of 220km/s.  While
this is certainly much smaller than its current galactocentric
distance, a small value was also found (using a similar argument and
published values of the mass and tidal radius) by \citet{vdb1995}.
The orbit given by \citet{Di1999} has a still smaller perigalactic
distance, the galactocentric distance varying between extremes of 0.6
and 5.9 kpc.



\subsubsection{Velocity dispersion profile}

This is illustrated in Fig.\ref{fig:vdp}, where it is compared with
the observational data of \citet{Pe1995}.  The shortfall at large
radii is of doubtful significance.

\subsubsection{Luminosity Functions}

These are shown for our model at the median radius of the innermost
and outermost fields observed by \citet{Ri2004}.  The disagreement in
the inner luminosity function at faint magnitudes may be attributable
to the fact that the theoretical result assumes 100\% completeness,
while the observational data are uncorrected for completeness.  A plot
of the completeness correction in the outer field is given by
\citet[Fig.3]{Ha2002} and discussed above in
Sec.\ref{sec:data_computation}.  The mismatch between model and data
is smaller in the outer field, but from the discussion above it would
seem that the mismatch in the faintest bin may be too large to be
accounted for by the estimated value of the completeness correction.
The error bars in the last two observational points on this plot (not
shown) almost overlap, and much of the mismatch may be simply sampling
uncertainty.  It is worth comparing the multi-mass King model
constructed by \citet{Ri2004}, which is also problematic in the
outermost field.



\subsubsection{Core collapse}

\begin{figure}
{\includegraphics[height=10cm,angle=-90,width=8cm]{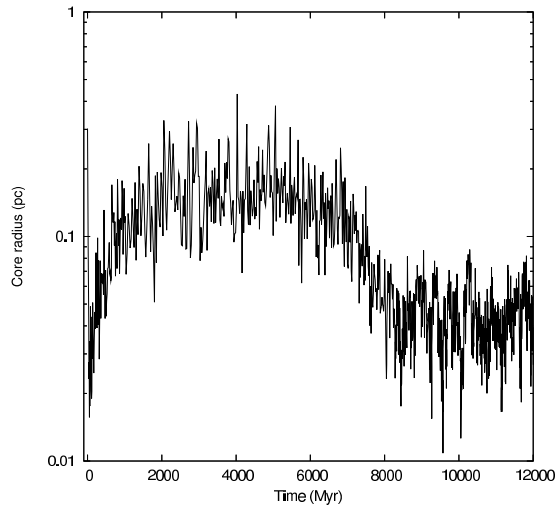}}
    \caption{Theoretical core radius of our Monte Carlo model.}\label{fig:rc}
  \end{figure}

Fig.\ref{fig:rc} shows the evolution with time of the theoretical core
radius.  There is an early period of very rapid contraction,
associated with mass segregation, followed by a slower reexpansion,
caused by the loss of mass from the evolving massive stars which are
now concentrated within the core.  

Our most surprising discovery from
our model of M4 is the subsequent behaviour.
M4 is classified as a King-profile cluster \citep{Tr1993}, and such
clusters are usually interpreted as being clusters whose cores  have
not yet collapsed.  But the plot of the theoretical core radius
(Fig.\ref{fig:rc} reveals that the model
 exhibited core collapse at about 8Gyr.  Subsequently
its core radius is presumably sustained by binary burning.  Even
non-primordial binaries may be playing a role here.  To the
best of our knowledge it has not previously been suggested that M4, is
a post-collapse cluster, though on statistical grounds
\citet{dempp2007} have suggested that some King-type clusters have
already collapsed.  This issue is often approached by reference to the
half-mass relaxation time (Table \ref{tab:m4dat}, but for M4 this is not short
enough to be decisive.

The presence of radial colour gradients is correlated with the
presence of a non-King surface brightness profile and hence with core
collapse.  We have computed the colour profile of our model, and find
that it is nearly flat except for the influence of one or two very
bright stars within the innermost few arcsec.  Most values lie around
0.65 in $B-V$, which is much less than the global value of 1.03 given
in \citet[January 2008]{Ha1996}.  On the other hand roughly similar mismatches
between observations are found in other clusters (e.g. M30, \citep{Pi1988}).


\subsubsection{The colour-magnitude diagram}

\begin{figure}
\begin{center}
\includegraphics[angle=-90,width=13cm]{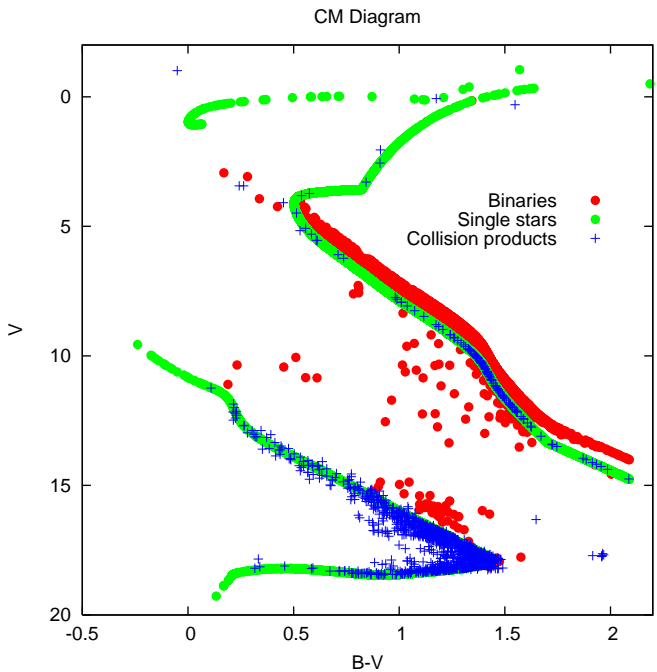}
     \caption{The colour-magnitude diagram at 12Gyr.  Green: single
 stars; red: binaries; blue pluses: collision or merger remnants.}
 \label{fig:colour-magnitude}
\end{center}
   \end{figure}

Figure \ref{fig:colour-magnitude} shows the colour-magnitude diagram of the model.
This is of interest, not so much for comparison with observations, but
for the presence of a number of interesting features.  The division of
the lower main sequence is simply an artifact of the way binary masses
were selected (a total mass above 0.2$\msun$ and a component mass
above 0.1$\msun$.)  Of particular interest are the high numbers of
merger remnants on the lower white dwarf sequence.  There are very few
blue stragglers.   Partly this is a result of the low binary
frequency, but it is also
important to note that some formation channels are unrepresented in
our models (in particular, collisions during triple or four-body
interactions, though if a binary emerges from an interaction with
appropriate parameters, it will be treated as merged.)  These numbers
also depend on the assumed initial distribution of semi-major axis,
which is not yet well constrained by observations in globular clusters.

Note in Fig.\ref{fig:colour-magnitude} that there are some white dwarf-main sequence binaries (below the
main sequence).  \citet{Ri2004} drew attention to the possible
presence of such binaries in their colour magnitude diagrams of this
cluster.

\subsubsection{Binaries}

\begin{figure}
    \begin{minipage}[]{0.45\textwidth}
{\includegraphics[height=10cm,angle=-90,width=8cm]{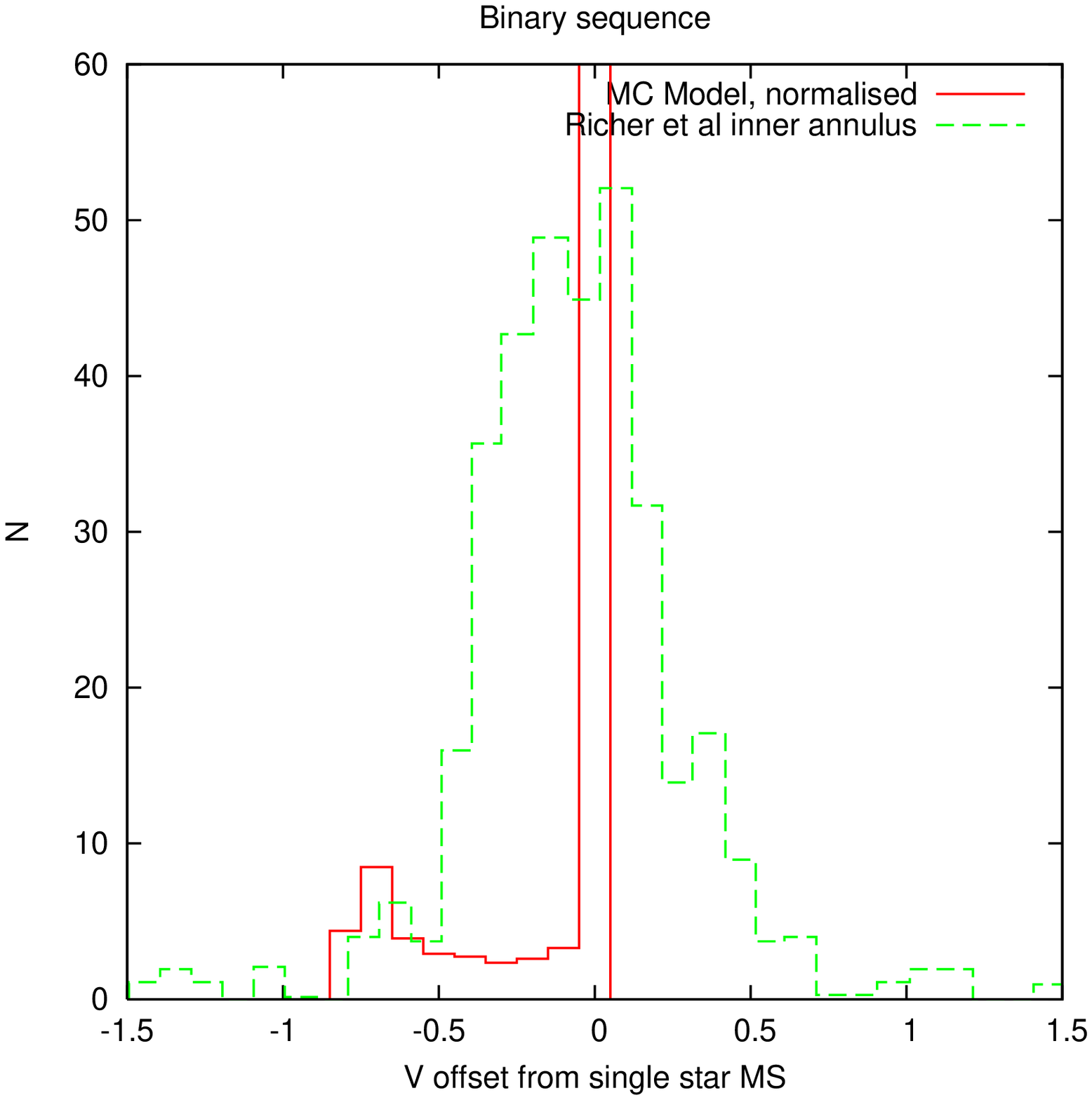}}
    \caption{Histogram of $V$-offset from the main sequence, compared
with the corresponding data from the innermost annulus studied by
Richer et al. (2004) See text for details.}
\label{fig:photometrics}
    \end{minipage}
\hfill
\begin{minipage}[]{0.45\textwidth}
{\includegraphics[height=10cm,angle=-90,width=8cm]{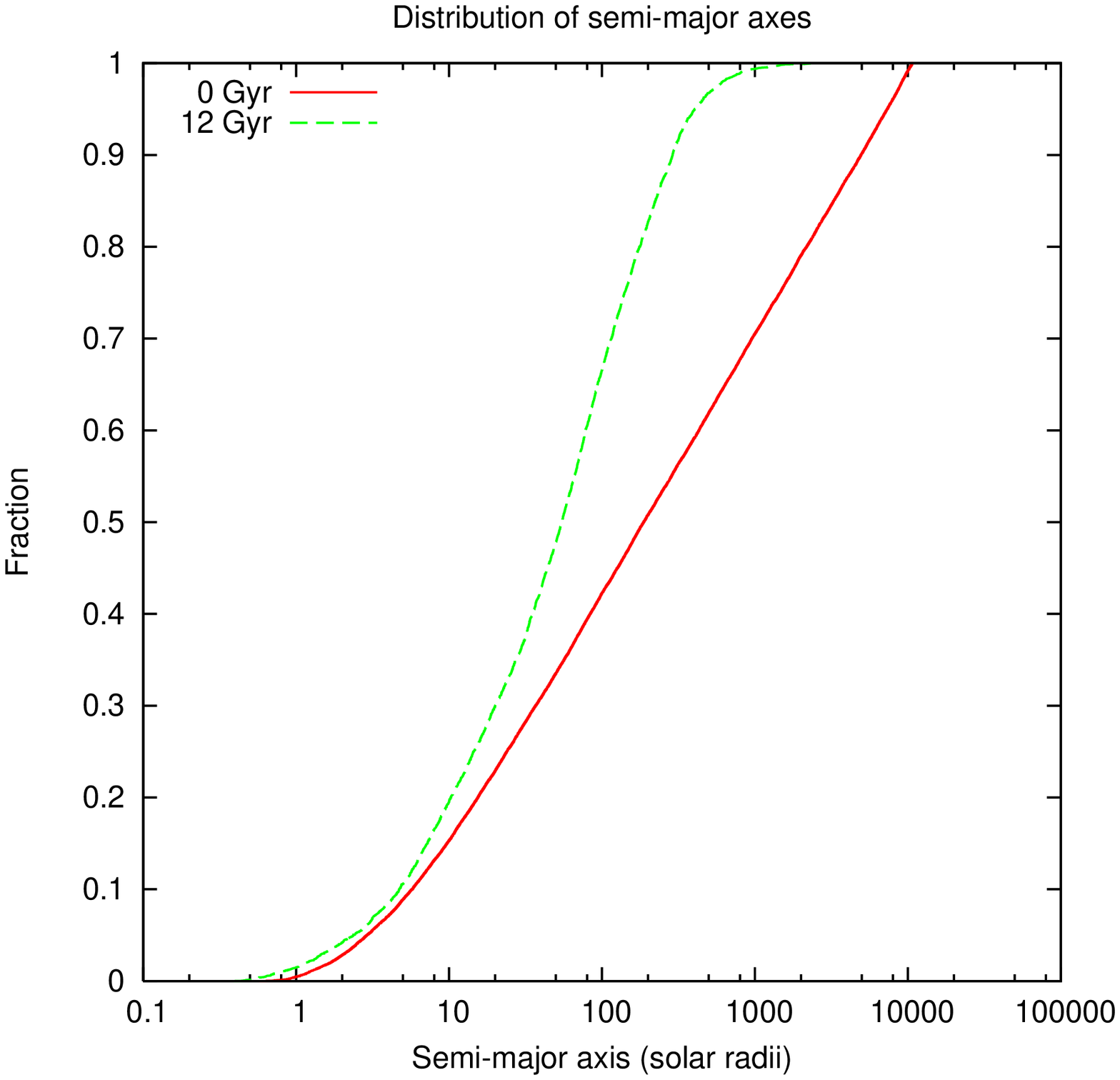}}
\caption{Distribution function of the semi-major axes of the
binaries at 0Gyr and 12Gyr.  Units: solar radii.}
\label{fig:sma}
\end{minipage}
  \end{figure}

Photometric binaries are visible in Fig.\ref{fig:colour-magnitude},
and these are compared with observations in the inner field of
\citet{Ri2004} in Fig.\ref{fig:photometrics}.  In this figure, the
model histogram has been normalised to the same total number of stars
as the observational one.  We made no attempt to simulate photometric
errors, but the bins around abscissa = -0.75 suggest that the binary
fractions in the model (which is under 6\% globally; see Table
\ref{tab:mc_king}) and the observations are comparable.  We note here
that \citet{Ri2004}, using this same data, concluded that the binary
frequency was approximately 2\% in the innermost field.  But they also
note that the measured frequency of ``approximately equal-mass
binaries'' is 2.2\%, and we consider that the balance between this
number and our binary fraction can be made of binaries with companions
whose masses are more unequal.

Radial velocity binaries should also be detectable in M4, and will be
part of a separate investigation.

Now we consider the distributions of the binaries, as predicted by the
theoretical model.  
  Fig.\ref{fig:sma} shows that binaries have evolved dynamically as well as through their internal
evolution.  In particular the softest pairs been
almost destroyed.

\begin{figure}
  \begin{minipage}[]{0.45\textwidth}
{\includegraphics[height=10cm,angle=-90,width=8cm]{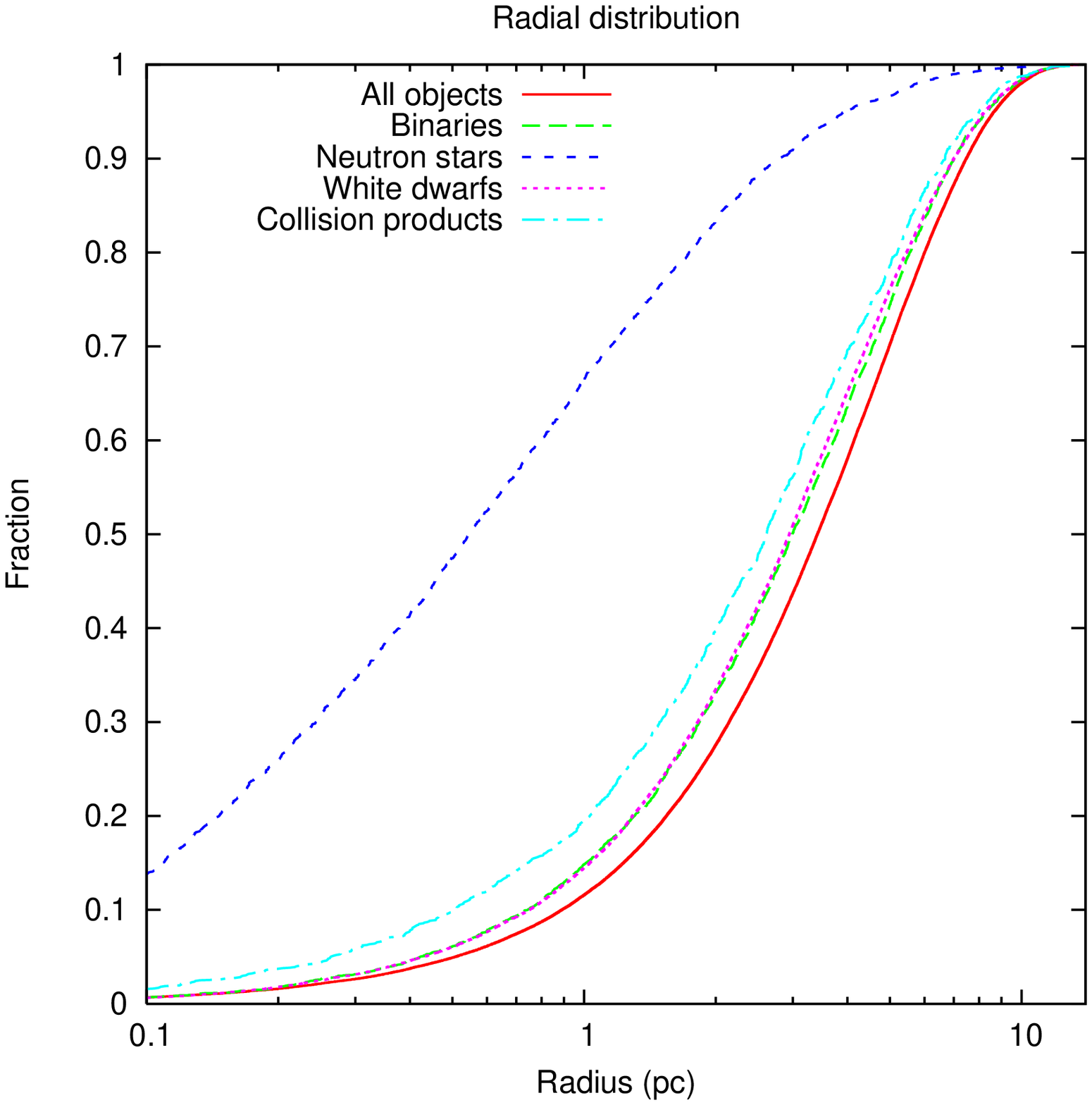}}
\caption{Radial distribution functions at 12Gyr.  Units: parsecs.  The
key identifies the class of object included.  The distributions of
white dwarfs and binaries are almost identical.}
\label{fig:segregation}
  \end{minipage}
\hfill
\begin{minipage}[]{0.45\textwidth}
{\includegraphics[height=10cm,angle=-90,width=8cm]{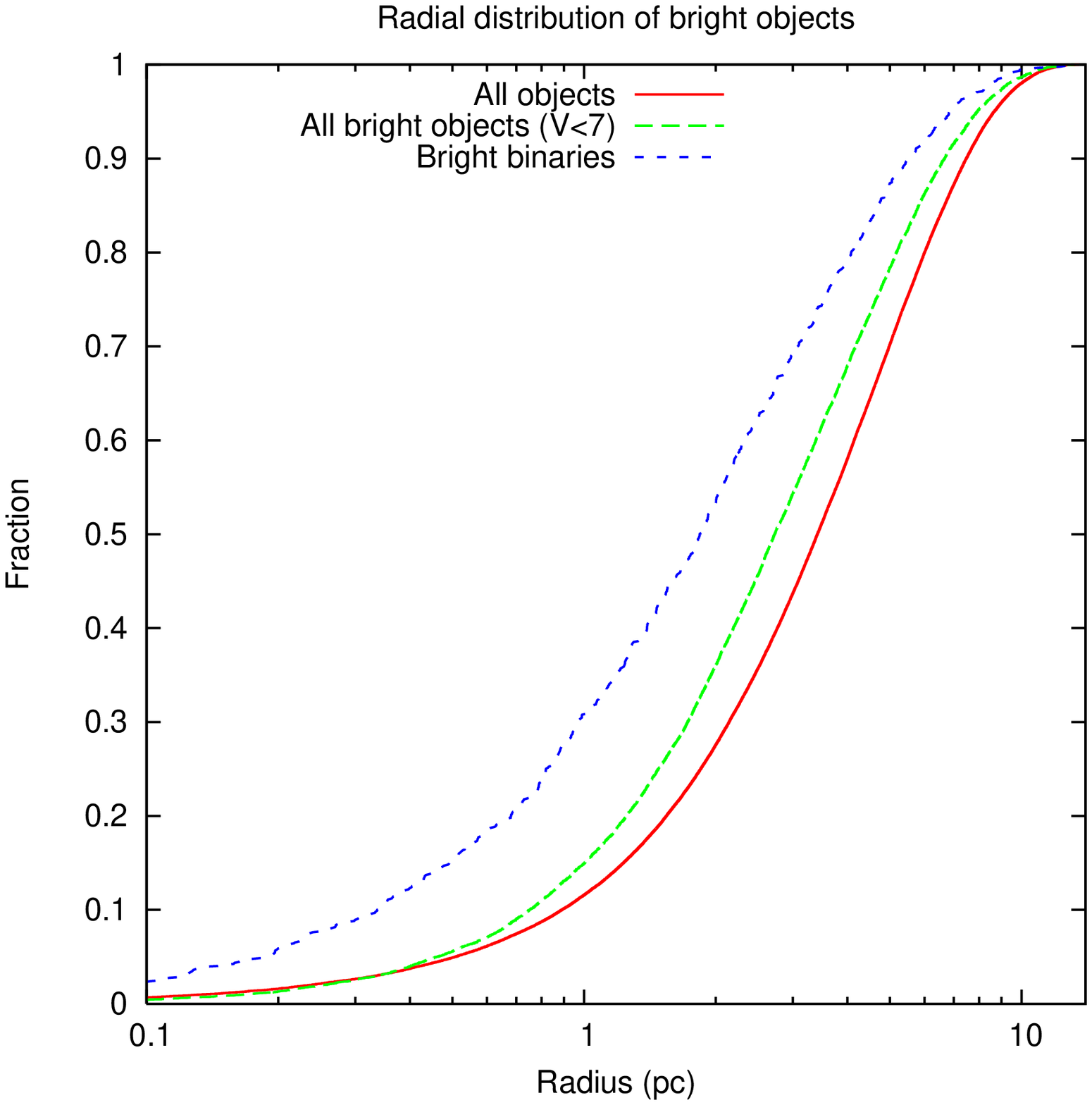}}
\caption{Radial distribution functions at 12Gyr, showing the extent of
segregation between bright single and bright binary stars
($M_V<7$). Also shown for comparison is the distribution for all
objects, as in Fig.\ref{fig:segregation}.  Units: parsecs.}
\label{fig:brights}
\end{minipage}
\end{figure}

By 12 Gyr the binaries exhibit segregation towards the centre of the
cluster, but perhaps in more subtle ways than might be expected
(Figs.\ref{fig:segregation},\ref{fig:brights}).  When {\sl all}
binaries are considered, there is little segregation relative to the
other objects in the system.  (Most binaries in our model are of low mass.)  But if one
restricts attention to bright binaries, which we here take to mean
those with $M_V < 7$ (i.e. brighter than about two magnitudes below
turnoff), the segregation is very noticeable (Fig.\ref{fig:brights}),
with a half-mass radius smaller by almost a factor of 2 than for bright
single stars.  Still, bright binaries are not nearly as
mass-segregated as neutron stars (Fig.\ref{fig:segregation}), which,
incidentally, receive no natal kicks in our model.  

The history of the binary fraction is, in effect, given in
Fig.\ref{fig:degenerate-numbers}.  In the first Gyr this falls
steadily from the initial value
of 0.07 to about 0.06, and it remains close to this value until the
present day.

\subsubsection{Escape velocity}

\begin{figure}
{\includegraphics[height=10cm,angle=-90,width=8cm]{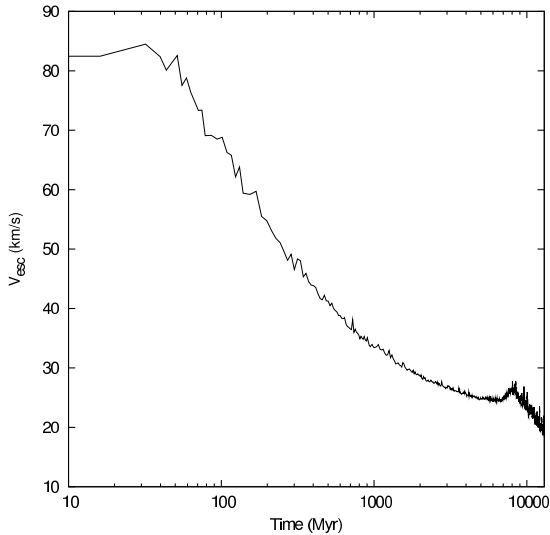}}
    \caption{Central escape velocity as a function of time.}
\label{fig:vesc}
  \end{figure}

We have already referred to the fact that, in our model, neutron stars
receive no natal kicks.  Escape is of course governed by the escape
speed, and this is of much interest in connection with the possibility
of retaining gas from the first generation of rapidly evolving stars.
For this reason we plot the central escape velocity in
Fig. \ref{fig:vesc}.  The remarkably high value in the first few tens
of millions of years, if valid for M4, could have interesting
consequences for the early evolution of the cluster and its stars.  It
draws attention to the very high initial density of our model (Table
\ref{tab:mc_king}), whose average value within the half-mass radius is
$2\times10^5\msun$/pc$^3$.  The central density is about
$10^6\msun$/pc$^3$, about an order of magnitude larger than in the
central young cluster in the HII region NGC
3603 \citep{St2006}.  In the absence of local young star clusters as
massive as our M4 progenitor, it is difficult to be sure whether our
initial model is implausibly dense.

\subsubsection{Degenerate components}

We have already mentioned the spatial segregation of the population of
neutron stars, and the presence of a number of degenerate binaries at
the present day.  Now we consider the historical evolution of this
population over the lifetime of the cluster so far, according to our
model.

  \begin{figure}
{\includegraphics[height=10cm,angle=-90,width=8cm]{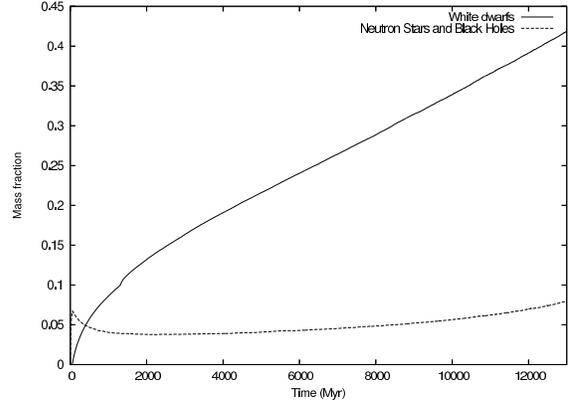}}
    \caption{Mass of white dwarfs, and of neutron stars and black
    holes together.  }
\label{fig:degenerate-mass}
  \end{figure}

Fig.\ref{fig:degenerate-mass} shows the evolution of these
populations.  It has already become well established
\citep{VH1997,HS2003} that white dwarfs account for an increasing
proportion of the mass of globular clusters, and indeed it is of order
39\% at the present day, according to this model.  The proportion of
neutron stars is almost certainly excessive, because of our
assumption of complete retention.

\begin{figure}
{\includegraphics[height=10cm,angle=-90,width=8cm]{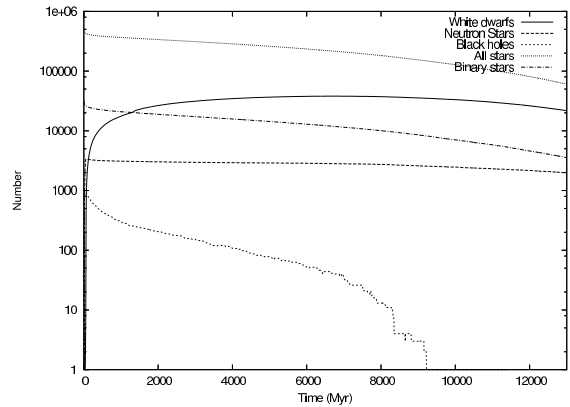}}
    \caption{Numbers of all stars,  white dwarfs,  neutron stars and black
    holes.  }
\label{fig:degenerate-numbers}
  \end{figure}

In order to separate  two of the  components in this figure (black
holes and neutron stars), we show in Fig. \ref{fig:degenerate-numbers}
the numbers of these degenerate stars, along with the numbers of all
stars and all binaries.  This shows that no stellar-mass black holes
are expected to be present in M4 now.

\section{Conclusions and discussion}

In this paper we have presented a Monte Carlo model for the nearby
globular cluster M4.  This model includes the effects of two-body
relaxation, evaporation across the tidal boundary, dynamical
interactions involving primordial and three-body binaries, and the
internal evolution of single stars and binaries.  By adjustment of the
initial parameters (total mass, tidal radius, initial mass function,
binary fraction) we have found a model which, after 12Gyr of
evolution, leads to a model with a surface brightness profile,
velocity dispersion profile, and luminosity functions at two radii,
all of which are in tolerable agreement with observational data.  It
also leads to a number of photometric binaries roughly consistent with
observation. 

This model has a current mass of $4.6\times10^4\msun$, a $M/L_V$ ratio
of 1.8, a binary fraction of almost 6\%, and an almost flat lower mass
function (Table \ref{tab:mc_king}).  Almost 40\% of the mass is in
white dwarfs, and about 7\% in neutron stars, though our model assumes
a 100\% retention rate.  They are strongly mass-segregated to the
centre of the cluster.  There are no stellar-mass black holes left in
the cluster.  The binaries have experience significant dynamical
evolution, almost all the soft pairs having been destroyed.  The
binary population as a whole is only slightly segregated towards the
centre, but there is more evident segregation of bright binaries (such
as those that would be more readily observed in a radial velocity
search.) 

The most significant new result from our model is the implication
that M4 is a core collapse cluster, despite the uncollapsed appearance
of the surface brightness profile.

Now we discuss a number of shortcomings of our model and other issues
related to this study.
\begin{enumerate}
  \item {\sl Uniqueness:}  Though we have arrived at a broadly
  satisfactory model, it is not at all clear how unique it is.
  Certainly we were unable to construct models with much larger
  numbers of primordial binaries, or with significantly larger initial
  radius, as such models produced an insufficiently concentrated
  surface brightness profile.  Furthermore, the fact that the lower
  slope of the initial mass function is less steep than the canonical
  value of 1.3 should not be taken to imply that we have inferred the
  initial value.
\item {\sl Fluctuations}: we have noticed that runs differing only in
  the initial seed of the random number generator can give
  surprisingly different surface brightness profiles.  In broad terms
  this confirms the important finding of \citet{Hu2007}, though we
  have not yet established (as he did) that the presence or absence of
  black hole binaries is the underlying mechanism.  We shall return to
  this issue at appropriate length in the next paper in this series,
  on the cluster NGC 6397.
\item {\sl The Initial Model:}  The initial model is astonishingly
  dense, the central density being of order $10^6\msun$pc$^{-3}$.  It
  also has to be realised that we are imposing initial conditions at a
  time when the residual gas from the birth of the cluster has already
  dispersed.  Various properties of a star cluster, including
  its binary population and mass function, may change significantly
  during the phase of gas expulsion, which further undermines the
  power of our model to establish the initial conditions.  
  Finally, we have assumed no initial mass segregation, even though
  recent work suggests that this should be present already before the
  cluster has assembled into a roughly spherical object
  \citep{Ve2007}.  This is an aspect of the modelling that could be readily
  improved.
\item {\sl The early evolution of the model:}  It is worth drawing
  attention here to the high initial escape velocity from the centre
  of our model, in the first few tens of millions of years.  Clearly
  this is dependent on the small initial radius.  The initial high
  density is also responsible for the very rapid initial evolution of
  the core.  
\item {\sl Imperfections of the modelling:} Several important
  improvements need to be made to our technique.
  \begin{enumerate}
\item   At present few-body
  interactions are handled with cross sections.  The problem is not
  simply that these are not well known, especially in the case of unequal
  masses; it also means that we are unable to determine if a collision
  occurs during a long-lived interaction.  For this reason, we have
  said almost nothing about collision products (e.g. blue stragglers)
  in this paper.  This
limitation could be overcome by direct integration of the
interactions, as is done by \citet{FR2007} in their version of the
Monte Carlo scheme. 
\item Long-lived triples  are neglected in the model at present.  These are commonly produced in
binary-binary encounters \citep{Mi1984a}, and it is desirable to
include these as a third species (beyond single and binary stars).
Their observable effects may be small, but of course there is one
intriguing example in the very cluster we have focused on here \citep{Th1993}.
\item We assume that the tide is modelled as a tidal cutoff.  We have
  taken some care to ensure that the tidal radius is adjusted so that
  the model loses mass through escape at the same rate as an $N$-body
  model would, if immersed in a tidal {\sl field} with the same tidal
  radius.  This means, however, that the effective tidal radius of the
  model is somewhat too small.  A better treatment of a steady tide
  may be possible, without this drawback.
\item We assume that the tide is steady, something which is not true
  for M4.    The effects of tidal shocks have
been studied by a number of authors \citep[e.g.][]{Ku1995}, and
it would be possible to add the effects as another process altering
the energies and angular momenta of the stars in the simulation.   On the other hand \citet{Ag1988} found, on the basis of a
  simple model, that tidal evaporation was the dominant mechanism in
  the evolution of the mass of M4 at the present day, and that other
  factors (such as disk and bulge shocking) contributed at a level
  less than 1\%.
\item Rotation: it has been shown \citep{Ki2004} that, to the extent
that rotating and non-rotating models can be compared, rotation
somewhat accelerates the rate of core collapse.  Rotation is hard to
implement in this Monte Carlo model, however.
\item The search for initial conditions is still very laborious, and
  we are constantly seeking ways of expediting this.  Our most
  fruitful technique at present is the use of small-scale models which
  relax at the same rate as a full-sized model.  Not all aspects of
  the dynamics are faithfully rendered by a scaled model, but the
  technique appears to be successful as long as the fraction of
  primordial binaries is not too large.
\end{enumerate}
  \end{enumerate}

Despite these caveats and shortcomings, it is clear that the Monte
Carlo code we have been developing is now an extraordinarily useful
tool for assessing the dynamics of rich star clusters.  For the
cluster M4 it has led to the conclusion that M4 can be explained as a
post-collapse clusters.  It also provides a wealth of information on
the distribution of the binary population, which will be important for
the planning and interpretation of searches for radial velocity
binaries, now under way. 

$N$-body models will eventually supplant the
Monte Carlo technique, but at present are incapable of providing a
star-by-star model of even such a small cluster as M4.  They can of
course provide a great deal of general  guidance on the dynamical
evolution of rich star clusters, and they underpin the Monte Carlo by
providing benchmarks for small models.  Even when $N$-body
simulations eventually become possible, Monte Carlo models will remain as a quicker
way of exploring the main issues, just as King models have continued
to dominate the field of star cluster modelling even when more
advanced methods (e.g. Fokker-Planck models) have become available.

\section{Discussion}

\section*{Acknowledgements}

We are indebted to Jarrod Hurley for much help with the BSE stellar
evolution package. We thank Janusz Kaluzny for his kind advice
on observational matters, and Harvey Richer for comments on various
matters, especially the observed luminosity functions. This research was 
supported in part by the Polish National Committee for Scientific 
Research under grant 1 P03D 002 27.

\bsp

\label{lastpage}

\end{document}